\documentclass[showpacs,amsmath,amssymb,floatfix,pre,twocolumn,superscriptaddress]{revtex4} 
\usepackage[dvips]{graphicx} 
\usepackage[latin1]{inputenc}
\usepackage{color,rotating}
\usepackage{amssymb} 
\usepackage{amsmath} 
\usepackage{martin_specials}
\usepackage{pilT_project} 
\usepackage[english]{babel}
\newcommand{\Ref}{Ref.\ }
\newcommand{\figpart}[1]{{(#1)}}
\date{\today}
\begin{document}
\title{Force generation in small ensembles of Brownian motors}

\author{Martin Lind\'en}
\email{linden@kth.se} 
\affiliation{Theoretical Physics, Royal Institute of Technology,
AlbaNova, 10691 Stockholm, Sweden}
\author{Tomi Tuohimaa}
\email{tomi.tuohimaa@biox.kth.se} 
\affiliation{Applied Physics, Royal Institute of Technology,
AlbaNova, 10691 Stockholm, Sweden}

\author{Ann-Beth Jonsson} 
\email{Ann-Beth.Jonsson@imbim.uu.se}
\affiliation{Department of Medical Biochemistry and Microbiology,
Uppsala Biomedical Center, Uppsala University, Box 582, 75123 Uppsala,
Sweden} 

\author{Mats Wallin} \email{wallin@kth.se} \affiliation{Theoretical
Physics, Royal Institute of Technology, AlbaNova, 10691 Stockholm,
Sweden}

\begin{abstract}
The motility of certain gram-negative bacteria is mediated by
retraction of type IV pili surface filaments, which are essential for
infectivity.
The retraction is powered by a strong molecular motor protein, PilT,
producing very high forces that can exceed 150 pN.
The molecular details of the motor mechanism are still largely
unknown, while other features have been identified, such as the
ring-shaped protein structure of the PilT motor.
The surprisingly high forces generated by the PilT system motivate a
model investigation of the generation of large forces in molecular
motors. We propose a simple model, involving a small ensemble of motor
subunits interacting through the deformations on a circular backbone
with finite stiffness.
The model describes the motor subunits in terms of diffusing particles
in an asymmetric, time-dependent binding potential (flashing ratchet
potential), roughly corresponding to the ATP hydrolysis cycle.
We compute force-velocity relations in a subset of the parameter space
and explore how the maximum force (stall force) is determined by
stiffness, binding strength, ensemble size, and degree of asymmetry.
We identify two qualitatively different regimes of operation depending
on the relation between ensemble size and asymmetry. In the transition
between these two regimes, the stall force depends nonlinearly on the
number of motor subunits.
Compared to its constituents without interactions, we find higher
efficiency and qualitatively different force-velocity relations.
The model captures several of the qualitative features obtained in
experiments on pilus retraction forces, such as roughly constant
velocity at low applied forces and insensitivity in the stall force to
changes in the ATP concentration.
\end{abstract}





\maketitle

\section{Introduction}
Recent experimental progress has enabled remarkable quantitative
measurement of biological processes on the single-molecule level
\citep{sheetz98}.  One example is the biomechanics of force generation
by molecular machines such as kinesin, myosin, and dynein
\citep{boal,howard}.  This has stimulated considerable modeling
activity in order to analyze the experiments
\citep{howard,boal,reimann,bustamante01,schliwa03}.  In this paper we
are inspired by another motor protein, called PilT \citep{mattick},
which has interesting properties; e.g., it is the strongest known
molecular motor \citep{maier02}.

The PilT motor is responsible for the retraction of certain bacterial
surface filaments, and the velocities and forces generated during
retraction have been measured in a series of laser tweezers
experiments \citep{merz00,maier02,maier04}.  Theoretical analysis of
the retraction data has revealed interesting information about the
underlying retraction mechanism \citep{maier04}. It is also of
interest to study the question of how large the generated forces can
be, given the energy and length scales relevant to the PilT motor, and
what features are important for generation of large forces. We will
address this question through a simple ratchet model, which is
inspired by known experimental facts of the PilT system.

Ratchet models of particles in fluctuating potentials are commonly
used in theoretical studies of molecular motors
\citep{reimann,prost97b}. Single-particle models have been used to
study kinetics of ATP consumption in molecular motors
\citep{lattanzi01} and to describe the kinetics of kinesin
\citep{astumian94}. Models of particles in ratchet potentials have
also been employed to describe collective effects in large ensembles
of interacting motors
\citep{shu04,badoual02,lipowsky01,prost95,prost97,plischke99,juelicher97}.
Finite ensembles of Brownian particles have been studied to some
extent in the context of two-headed motor proteins
\citep{derenyi96,vilfan99,csahok,dan03,klumpp01} and to describe the
bacterial flagella motor \cite{xing06}.

Another approach, in which the motion of a molecular motor is
described in terms of transitions between discrete chemical and
conformational states, has also been generalized to the case of two
interacting motor subunits \citep{kolomeisky05}.  As is evident from
Refs.\
\citep{shu04,badoual02,lipowsky01,prost95,prost97,plischke99,juelicher97,derenyi96,vilfan99,csahok,dan03,klumpp01,xing06},
interactions among several motors can lead to new and nontrivial
behavior of the average velocity, which is not present if the
interaction is turned off. It is natural to ask if this is the case
for force generation as well.

In this paper, we investigate the behavior of a small ensemble of
interacting processive Brownian motors and focus on the effect of the
interaction on the generation of large forces.  We aim at a
prototypical, minimal model which captures certain features of the
PilT system. The main input is the overall structures of the filament
and PilT complex. When possible, we also use the experimental
situation to estimate model parameters, which should ideally be as few
as possible.  The model is a generalization of the model of two
elastically coupled motors studied by \citet{dan03} to larger
ensembles, but focuses on different properties. By varying the density
of motor subunits and other parameters of the system, we explore the
force production in different regimes.  We compare with a single
building block of our model to identify the effect of the interactions
and also compare with experimental results.  Although the detailed
connection between the model and the actual molecular retraction
mechanism is speculative, the spirit of the model is best understood
in light of the known facts about the PilT system. Therefore, we will
briefly review some facts about pilus retraction before introducing
the model.

Type IV pili are surface filaments crucial for the initial adherence
of certain gram-negative bacteria to target host cells, DNA uptake,
cell signaling, and bacterial motility \citep{mattick}. Each filament
consists of thousands of pilin monomers that polymerize to a helical
structure with outer diameter of about 6 nm, 4 nm pitch, and five
monomers per turn \citep{mattick,forest97}. The bacterial motility
associated with type IV pili, called twitching motility, is driven by
repeated extension, tip attachment, and retraction of the pilus
filament, by which the bacterium can pull itself forward on surfaces
like glass plates or target host cells \citep{merz00}. Type IV pili
are expressed by a wide range of gram-negative bacteria
\citep{mattick} including \bakterie{Myxococcus xanthus} \citep{sun}
and human pathogens \bakterie{Neisseria gonorrhoeae} \citep{merz00}
and \bakterie{Pseudomonas aeruginosa} \citep{skerker}.

The mechanism of retraction is believed to be filament disassembly
mediated by PilT, a member of the AAA family of motor proteins
\citep{mattick}, but the microscopic details of this process are not
known.  One might compare pilus retraction with force generation by
microtubules, which are multistranded filaments with effective monomer
lengths similar to type IV pili. The helical pitch divided by the
number of filament strands is 0.8 nm for the pili and 0.6 nm for
microtubules \citep{howard}. However, pilus retraction generates
forces of up to \mbox{160 pN} \citep{maier02,maier04}, which is an
order of magnitude larger than those observed in \emph{in vitro}
experiments on microtubules \cite{howard,dogterom02,janson04}. Another
difference is that the pilus retraction velocity is independent of
filament length \citep{maier02}. Since dissociated pilin monomers are
stored in the cell membrane waiting to be recycled in other filaments
\cite{skerker}, the implication is that the velocity is independent of
pilin concentration in the membrane. This presumably rules out simple
polymerization ratchet-type models, which have been proposed to
describe polymerization forces generated by microtubules
\citep{howard,dogterom02}. The experimental evidence instead favors a
retraction process driven by an active molecular motor
\citep{maier02}.

Pilus retraction is highly processive, and retraction velocities are
of the order of $0.5-1$ $\mathrm{\mu}$m/s
\citep{mattick,skerker,maier02,maier04}.  Generation of high forces
persists when the PilT concentration is reduced, suggesting that one
single PilT complex retracts the pilus filament \citep{maier02}. The
stall force (the force at which the average velocity drops to zero)
and the velocity at high forces are insensitive to changes in ATP
concentration, and the retraction velocity is roughly
force-independent for small applied forces ($\lesssim 50$ pN) within
the experimental accuracy \citep{maier02,maier04}. PilT has been shown
to form a ring structure with sixfold symmetry \citep{forest04}, and
since each subunit has an ATP binding motif, it is possible that it
can hydrolyze up to six ATP molecules in parallel during retraction
\citep{mattick}. The outer diameter of the ring is about 10 nm, and
the inner diameter varies in the range $2-4$ nm \citep{forest04}.

Pilus retraction is interesting from a technological point of view, as
a potential prototype for a nanomachine that can generate large
forces, and from a biomedical point of view since pilus retraction is
important for the infectivity of various severe bacterial pathogens
\citep{steroids}.

There are several proposals for how the molecular constituents of the
retraction machinery fit together. One of them is that PilT forms a
ring around the base of the pilus \citep{mattick,kaiser00}. The hole
in the middle of the PilT complex seems too small to let the assembled
filament through, but large enough for pilin monomers.  This could
allow interactions between the pilus and PilT via several active sites
(motor subunits) that work together and is the principle that we will
explore here. For simplicity, we assume one motor subunit per filament
strand and neglect possible two-dimensional effects such as angular
motion of the filament.

We stress that the purpose of this paper is not to attempt to describe
the detailed molecular mechanisms involved in pilus retraction, which
are largely unknown.  Rather we examine a new regime of a simple
model, whose main features are inspired by experiments. Below we
obtain several results from the model, such as large force generation
and other properties that agree well with interesting experimental
results on pilus retraction.  Moreover, these results are a
consequence of correlations and interactions between the motor
subunits and are strikingly different from the characteristics of the
single building block of the model.

\section{Retraction model}\label{sec_model}
In this section the geometry and equations of motion of the model are
described. We then discuss the parameters, which come in several
kinds: parameters that are known for the PilT system, parameters that
can be estimated to varying degrees of accuracy, and parameters that
we will explore in a systematic way.  A few parameters cannot be
estimated due to the lack of knowledge of the molecular details. In
this case, we make an arbitrary choice in order to investigate the
qualitative behavior of the model.
\begin{figure}[t!]\begin{center}
\includegraphics{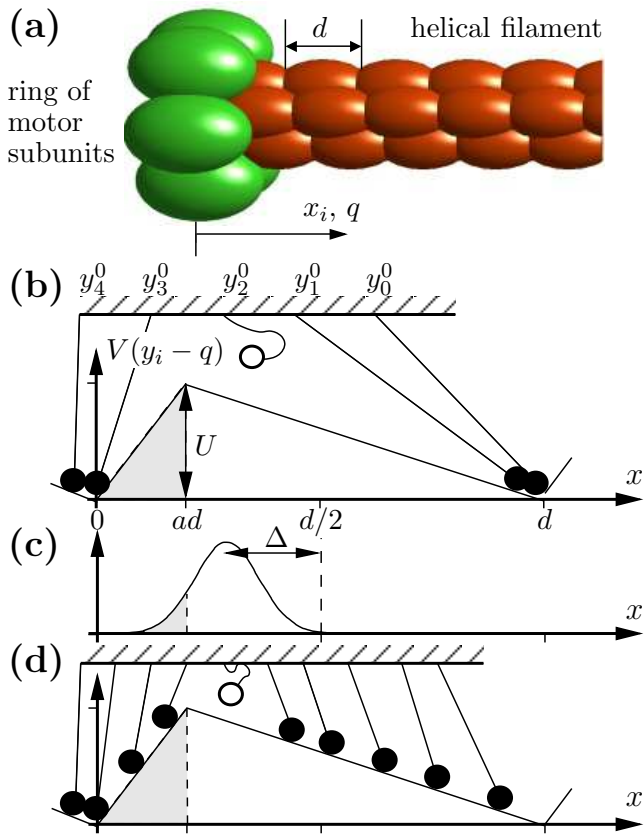}
\caption{\label{potential_stroke} (Color online) \figpart{a} The
elements of the retraction motor model consist of a flexible ring of
motor subunits that interact with a moving helical
filament. \figpart{b} Equivalent geometry after the change of
variables $y_i=x_i-id/M$, which places the binding potentials of the
filament monomers on top of each other.  The binding potential is
assumed to be an asymmetric ratchet potential with amplitude $U$ and
asymmetry $a$. The undeformed state of the motor protein complex is
described by the equilibrium positions $y_i^0$ of the motor subunits,
and the subunits are elastically confined to their equilibrium
positions. Due to the helical structure of the filament, the
equilibrium positions become evenly spread over one period. The motor
subunits at positions $y_i$ (black circles) interact with the filament
potential. The open circle represents an unbound subunit.  During a
successful retraction process, a motor subunit detaches from the
filament, relaxes in its confinement potential, and rebinds near the
next binding site along the filament.  \figpart{c} Distribution of the
unbound subunit in B. The shaded area represents the probability for
the subunit to bind to the left of the potential maximum and produce a
failed step. As the applied force increases, the filament is pulled
forward relative to the distribution and the probability of a failed
step increases. \figpart{d} Qualitatively new behavior emerges in the
limit of high stiffness (strong confinement) and many motor
subunits. In this limit, all bound motor subunits will not relax to
the minima in the binding potential. Instead, some subunits will
interact with the shaded part of the potential and oppose force
production. In this regime, one expects that the stall force depends
nonlinearly on the number of motor subunits, and nonmonotonically on
the stiffness.}\end{center}\end{figure}
The basic setup is sketched in Fig.\ \ref{potential_stroke}(a). A ring
of $M$ motor subunits interacts with an $M$-stranded helical filament,
with repeat distance $d$ of the single strands. The filament
coordinate $q$ decreases during retraction. We also allow for
deformations of the PilT ring. This is described by displacements
$x_i$ of the motor subunits from the undeformed state. We assume that
the motor subunits interact with the filament strands via identical
one-dimensional binding potentials with period $d$. To mimic the
helical structure of the filament, the potential of subunit $i$ is
displaced a distance $id/M$ relative to the potential of subunit $0$.
The interactions between the filament and subunit $i$ therefore have
the form $V(x_i-q-id/M)$, where $V$ is some one-dimensional binding
potential. We formulate the equations of motion as a system of
overdamped Langevin equations for $q$ and $x_i$,
\ekv{eom}{
\begin{split}
  \alpha \dot{x}_i &= -\km x_i - h_i(t)\frac{\partial
  V(x_i-q-id/M)}{\partial x_i}+\sqrt{2\alpha \kBT}\,\xi_i(t),\\ \gamma
  \dot{q}& = \sum_{i=0}^{M-1}h_i(t) \frac{\partial
  V(x_i-q-id/M)}{\partial x_i}+ \Ftrap+\sqrt{2\gamma
  \kBT}\,\xi_\text{q}(t),
\end{split}}
where $i=0,\ldots,M-1$, $\alpha$ and $\gamma$ are friction
coefficients of the motor subunits and the filament respectively,
$\km$ is a spring constant describing the stiffness of the PilT ring,
$h_i(t)=0,1$ are chemical state variables, $k_{\mathrm B}$ is
Boltzmann's constant, $T$ is the temperature, $\Ftrap$ is an external
force acting on the filament, and $M$ is the number of motor subunits
and filament strands. Thermal fluctuations are included through
independent Gaussian white noise terms $\xi_i(t)$ and
$\xi_\text{q}(t)$, which obey
\mbox{$\mean{\xi_i(t)}=\mean{\xi_\text{q}(t)}=0$},
\mbox{$\mean{\xi_j(t)\xi_i(t')}=\delta_{ij}\delta(t-t')$}, and
\mbox{$\mean{\xi_\text{q}(t)\xi_\text{q}(t')}=\delta(t-t')$} and have
prefactors according to the fluctuation-dissipation theorem. The
temperature is set to $T$= \mbox{310 K}.

Before discussing the terms in the equations of motion in detail, we
transform the subunit coordinates to place the binding potentials on
top of each other. The new coordinates are convenient in order to
analyze the model and are close to previous works with similar models
\citep{dan03,klumpp01,csahok}. The transformed subunit coordinates are
$y_i=x_i-id/M+d$. An undeformed PilT ring is now described as
$y_i=y_i^0=d-id/M$, which we call the equilibrium positions of the
subunits. The transformed equations of motions are
\ekv{eomYQ}{
\begin{split}
  \alpha \dot{y}_i &=-\km (y_i-y_i^0) - h_i(t)\frac{\partial
  V(y_i-q)}{\partial y_i}+\sqrt{2\alpha \kBT}\,\xi_i(t)\\ \gamma
  \dot{q}& = \sum_{i=0}^{M-1}h_i(t)\frac{\partial V(y_i-q)}{\partial
  y_i}+\Ftrap+\sqrt{2\gamma \kBT}\,\xi_\text{q}(t)
\end{split}}
Note that the binding potentials are all on top of each other, as if
all motor subunits were interacting with just one single strand.  The
price for this convenience is that the equilibrium positions are
evenly distributed over one period of the potential.  The setup in the
transformed variables is sketched in Fig.\ \ref{potential_stroke}(b).

The motor subunits can be in one of two states: unbound ($h_i=0$), in
which they diffuse around their respective equilibrium positions, or
bound ($h_i=1$), in which they also interact with the filament through
the binding potential $V(y_i-q)$.  We treat the number of motor
subunits $M$ as a parameter, and we will present results for $M=2$,
$3$, $5$, $6$, and $12$.

For simplicity, we model the binding of the motor to the filament by
an asymmetric sawtooth potential, shown in Fig.\
\ref{potential_stroke}(c).  For the asymmetry factor we take $a=0.1$,
if not stated otherwise. Some asymmetry is needed to give a preferred
direction of motion, and the sawtooth potential was selected to give a
simple parametrization of the (unknown) real interaction potential.
One possible origin of an asymmetric binding potential is surface
charges on the head of the pilin monomer \citep{parge95} -- for
example as described in Ref.\ \citep{boal}. Another alternative is
that the asymmetry can be viewed as an effective description of some
asymmetry somewhere else in the system, e.g., in the direction of the
motor steps (power strokes). Based on the helical structure of the
pilus filament, we take the periodicity of the potential to be $d=4$
nm \citep{forest97,mattick}, which we use for all values of $M$.

The amplitude $\dE$ of the potential is the maximal energy required to
break the bond between the filament and the active site. Pilus
retraction is powered by hydrolysis of one or a few ATP per retracted
pilin monomer \citep{maier02}, which sets the energy scale for the
potential. Depending on conditions, the free energy yield from
hydrolysis of one ATP in a cell is about $80-100$ pN nm
\citep{howard}. The motor subunits are bound together to form the
motor complex, and we model their confinement with a harmonic
restoring force $-\km x_i=-\km(y_i-y_i^0)$. This linear approximation
is reasonable if the deformations $x_i=y_i-y_i^0$ are small, which
will be verified below.

For the binding processes $h_i(t)$, we restrict ourselves to a
sequential reaction scheme with $M$ chemical states.  We define state
$j$ as $h_k(t)=1-\delta_{kj}$; i.e., subunit $j$ is unbound and the
other subunits are bound.  The states are visited in ascending order,
and the (constant) transition rate from state $j$ to $j+1(\text{mod}
M)$ is $\lambda$. The bound subunits spend most of their time near a
minimum in the binding potential, and several geometrical
configurations are compatible with each chemical state. The main
pathway for efficient retraction in the model is that the subunits
take turns to hop forward to the next minimum as they release and
rebind to the filament (successful steps). When the retraction is less
efficient -- for example at large applied force -- the subunits
sometimes do not hop forward (failed step), which leads to geometrical
configurations outside the main pathway. During a successful step, a
motor subunit goes through the following sequence of events:
\begin{enumerate}
  \item[(i)] The subunit is released from a minimum in the binding
    potential.
  \item[(ii)] The released subunit relaxes to its equilibrium position. The
    filament relaxes in the opposite direction due to the forces from
    the other motor subunits.
  \item[(iii)] The subunit rebinds, close to the next minimum in the binding
    potential (otherwise the step fails). At the same time, the next
    motor subunit enters step (i).
  \item[(iv)] The subunit stays bound and pulls on the filament as the
    other $M-1$ subunits go through steps (i)--(iii). After each
    successful rebinding event, the filament retracts a distance
    $d/M$.
\end{enumerate}
This mechanism relies on the asymmetry of the potential and is similar
to the mechanisms studied earlier for two elastically coupled
particles \citep{dan03,klumpp01}.
We argue below that generation of strong forces
in this model relies on a binding process that always keeps several
motor subunits bound to the filament, but the binding order is less
important.

We now discuss the parameters of the model.  In the laser tweezers
experiments \citep{maier04,maier02,merz00}, the outer filament tip
binds to an external latex bead with diameter $1-2$ $\mu$m.  Using
Stokes law, $\gamma=6\pi\eta R$, the approximate viscosity
$\eta=10^{-8}$ pN s/nm$^2$ of the bulk solution surrounding the cell
(somewhere between $10^{-9}$ for water and $8\times10^{-7}$ for
glycerin seems reasonable) and $R=1$ $\mu$m, we get $\gamma \approx
2\times 10^{-4}$ pN s/nm for the bead. As a first approximation, we
neglect the elasticity and friction of the pilus filament itself. For
the internal friction coefficient, we use $\alpha = 0.5 \times
10^{-3}$ pN s/nm $\ll\km/\lambda$. This sets the time scale for
internal relaxation $\alpha/\km$ much smaller than the typical time
$(M-1)\lambda^{-1}$ between binding and release of individual motor
subunits and lets the motor subunits reach thermal equilibrium between
transitions. This is consistent with estimates of thermal relaxation
times over length scales on the order of 10 nm \citep{juelicher97},
which is the size of the PilT ring. Another time scale for internal
relaxation is given by the time to slide down to a potential minimum,
$\alpha d^2/\dE$, which we also keep smaller than $(M-1)\lambda^{-1}$.
We then expect the velocity to be proportional to $\lambda$, and we
will restrict ourselves to this quasistatic regime for two
reasons. First, this is the biologically relevant regime where we
expect the stall force to be independent of $\lambda$, which
corresponds to the experimental observation that the stall force is
independent of ATP concentration \citep{maier02}. Second, the exact
value of $\alpha$ is not critical for the results in this regime, and
since it is difficult to estimate $\alpha$ accurately, we can avoid
making our results depend strongly on an unknown parameter.

Having found useful values for the potential period $d$ and the drag
coefficients $\alpha_i$ and $\gamma$, we go on to investigate the
model behavior as a function of the remaining parameters
$a,M,\km,\dE$, and $\lambda$ and properties of the binding process.

\section{Results}
\subsection{Methods}
Retraction of the filament means that $q$ decreases, so it is natural
to study the retraction velocity $v=-dq/dt$. In the laser tweezers
experiment, the tip of the bead is held by a static laser trap which
is to good approximation a harmonic potential -- i.e.,
\mbox{$\Ftrap=-\ktrap q$}, with $\ktrap$ on the order of $0.1$ pN/nm.
Numerical solution of Eq.\ \eqref{eom} using a standard method, known
as the Milstein scheme \citep{kloedenplaten}, produces a deflection
trajectory similar to the experimental ones. We calculate the
retraction velocity by fitting a second-order polynomial to a small
time interval around a point $q=-F/\ktrap$ and take the velocity
$v(F)$ as the derivative of the polynomial. This is similar in spirit
to how the experimental data was analyzed
\citep{merz00,maier02,maier04}. The retracted distance $-q(t)$
increases from the initial value towards a steady-state, corresponding
to the maximal applied force (stall force), which we define as the
mean applied force in the steady state.  To check our simulation code,
we reproduced analytical results for the steady state current in a
flashing ratchet model \citep{parrondo98}.
\subsection{Stall force and force-velocity 
relation}\label{qualitativeArguments} The stall force and
force-velocity relation of the motor is determined by several
competing mechanisms, which we now describe qualitatively.

In the case of a few motor subunits, stalling of the retraction is
controlled by two different mechanisms. One comes from the finite
binding energy between the filament and the subunits. Due to the
simple shape of the potential, we can estimate an upper limit for the
stall force, using force$=\Delta V/\Delta q$. The maximum force that
each subunit can exert on the filament during retraction against an
opposing (positive) force is $\dE/(1-a)d$, so $M-1$ bound motor
subunits give an upper limit of $\dE(M-1)/d(1-a)$ for the stall force.
At finite temperatures the upper limit is not reached, since the motor
subunits diffuse and can pass between potential minima by thermal
excitation. We think of these thermally assisted transitions as
slipping events, and they occur more often when the subunits are far
from their equilibrium positions and experience a large confining
force. This mechanism tends to increase the stall force with increased
binding strength $\dE$.

The other mechanism has to do with the stiffness $k$ and the
probability for a step to fail. This probability depends on the
distribution of the filament position relative to the unbound subunit,
which is illustrated in Fig.\ \ref{potential_stroke}(c). The shaded
area represents the probability of a failed step.  Higher probability
of failure gives lower velocity. Increasing the applied force $F$ at
constant stiffness pulls the filament to the right, increases the
fraction of failed steps, and decreases the velocity. If all steps
fail, no retraction takes place. The width of the position
distribution of a subunit in the harmonic confinement is
$\sqrt{\kBT/\km}$, but the distribution in Fig.\
\ref{potential_stroke}(c) is broader, since the filament also
fluctuates. The average relative position $\Delta$ is roughly
proportional to $F/\km$. At very low stiffness, the distribution is
broad enough for some steps to fail without applied load, and the
average relative position varies strongly with applied force. This
gives a monotonically decreasing force-velocity relation and a low
stall force limited by the stiffness. At very high stiffness, the
distribution is narrow and the average relative position is less
sensitive to the applied force. Almost no steps fail without applied
load, and it takes some threshold force before steps start to fail
significantly. We get a force-velocity relation that is almost
independent of force at low forces, and the stall force is mainly
limited by slipping events.
\begin{figure}[t]\begin{center}
    \includegraphics{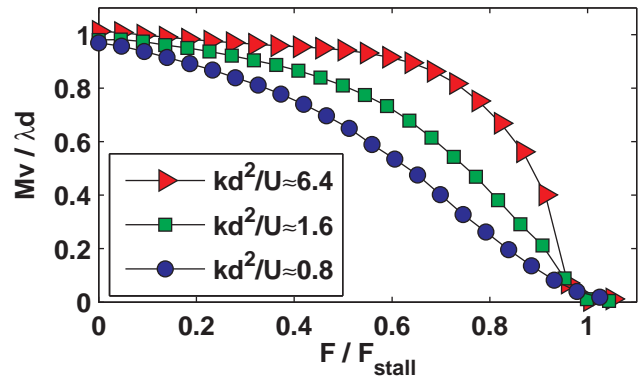}
    \caption{\label{vFrelations} (Color online) Force-velocity
    relations for high (triangles), intermediate (squares) and low
    stiffness (circles) calculated for $M=5$. The y axis is velocity
    normalized by $\lambda d/M$, which corresponds to the velocity for
    $M$ motor subunits if all steps were successful.}
\end{center}\end{figure}
Figure \ref{vFrelations} shows examples of normalized force-velocity
relations of the model for strong, intermediate, and weak stiffness,
compared to the binding strength. The curves illustrate the
qualitative arguments of the preceding paragraphs. We also find that
the stall force is insensitive to changes in the reaction rate
$\lambda$ (not shown). Since the reaction rate corresponds to the ATP
concentration, this is in qualitative agreement with the experimental
result that the stall force is insensitive to the ATP concentration
\citep{maier02}. This is expected in the quasistatic regime that
results from the choice of time scales discussed at the end of Sec.\
\ref{sec_model}. Stiff systems ($\km d^2/\dE\gg 1$) have a plateau in
the retraction velocity at low forces, which is also the general
experimental trend \citep{maier02,maier04}. For the parameter regime
we have investigated, Fig.\ \ref{vFrelations} gives the qualitative
shape of the force-velocity relation as a function of $\km/\dE$. The
size of the plateau is roughly proportional to the stall force, with a
proportionality constant that is to first approximation a function
only of $\km/\dE$.

As the number of motor subunits and filament strands increases, high
stiffness can also have a destructive effect on the stall force, as
illustrated in Fig.\ \ref{potential_stroke}D. This occurs when the
distance between equilibrium positions is shorter than the region of
the binding potential with backward slope (shaded) -- i.e.,
$M>1/a$. In that case, some of the bound motor subunits will tend to
interact with the shaded region of the potential if the stiffness is
high. There they act with a negative force on the filament and
contribute negatively to the force production. This effect is enhanced
by increased stiffness, and we therefore expect the stall force to
have a maximum as a function of stiffness. This is a qualitatively
different behavior than with only a few motor subunits ($M<1/a$) and
makes the stall force depend in a highly nonlinear way on $M$.

\subsection{Parametrization of the stall force}
\begin{figure}[t]\begin{center}
\includegraphics{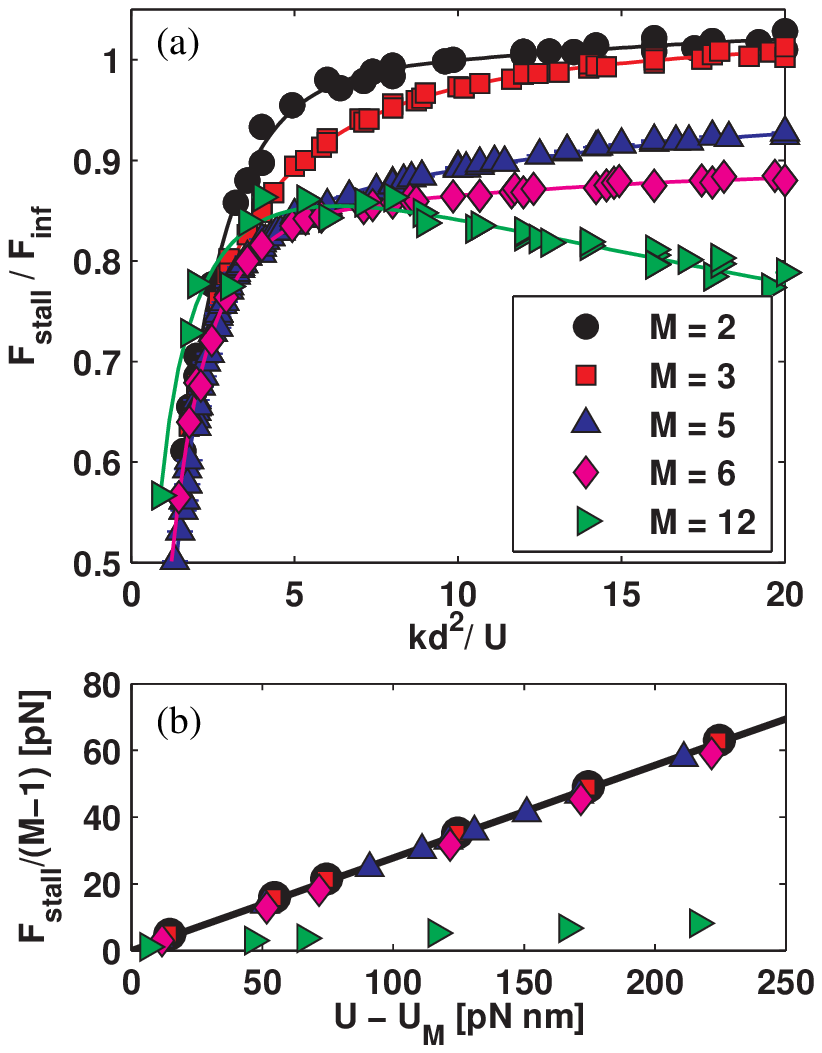}
\caption{\label{scaleFig} (Color online) \figpart{a} Stall force as a
function of $\km/\dE$ in the range \mbox{$5<\km<250$ pN/nm} and
\mbox{$80<\dE<200$ pN nm}. The normalization $\Finf$ is chosen
according to Eq.\ \eqref{scaleEq}. Lines are guides to the eye.  The
case of many subunits ($M=12$) is qualitatively different from the
other cases. Here, the stall force depends nonmonotonically on
$\km/\dE$. \figpart{b} Stall force per bound motor subunit in the
stiff limit $\km\to\infty$ as a function of $\dE-\dE_M$.  For $M\le 6$
the simulations are consistent with $f_M(\infty)=1$ in Eq.\
\eqref{scaleEq} (solid line). $M=12$ gives a lower stall force that
does not fit Eq.\ \eqref{scaleEq}, which again illustrates the
nonlinear behavior of the stall force on $M$.  Error bars in both
graphs are smaller than the symbols.}
\end{center}\end{figure}

Equation \eqref{eomYQ} suggests that the stall force might depend on
the ratio $\km /\dE$, instead of $\km$ and $\dE$ independently. We
will use this observation to further analyze force production in the
model.  A parametrization of the stall force is obtained from a
combination of the estimated upper limit $\dE(M-1)/d(1-a)$ for the
stall force with an function of $\km/\dE$.  Using this (unknown)
function $f_M$, which also depends on $M$, we describe the effect of
stiffness in the following way:
\ekv{scaleEq}{\begin{array}{cc}
\Fst(\km,\dE)=\Finf\,\fscale{M}{\frac{\km d^2}{\dE}},&
\Finf=\frac{M-1}{d(1-a)}\left(\dE -\dE_M\right),
  \end{array}}
where $\dE_M$ is a free $M$-dependent parameter, independent of $\dE$
and $\km$, and $d^2$ was inserted to make the argument of $f_M$
dimensionless.  One can obtain $\fscale{M}{x}$ by plotting
$\Fst/\Finf$ against $\km d^2/\dE$ and adjusting $\dE_M$. Such a plot
is shown in Fig.\ \ref{scaleFig}(a), with the best fit values of
$\dE_M=25.4$, $25.9$, $28.9$, $28.3$, and $33.5$ pN nm for $M$= 2,
3, 5, 6, and 12, respectively. For $M$=3, 5, and 6, the data points
fall on a single curve for each $M$ to good approximation, while $M$=2
and 12 show some scattering.
We can understand the ansatz \eqref{scaleEq} and the results in Fig.\
\ref{scaleFig} in light of the qualitative arguments in Sec.\
\ref{qualitativeArguments}.  We interpret $\Finf$ as the stall force
in the stiff limit $\km\to\infty$ and $\dE_M$ as the effect of thermal
fluctuations, which induce slipping events and thereby lower the
maximal force that the binding potential can support.  The stiff limit
$\km\to\infty$ can be simulated by locking the motor subunits to their
equilibrium positions. The results are shown in Fig.\
\ref{scaleFig}(b), using the above values for $\dE_M$.  The stall
force per bound motor subunit is consistent with Eq.\ \eqref{scaleEq}
and $\lim_{x\to\infty}\fscale{M}{x}=1$ for $M=2$, $3$, $5$, and
$6$.  The case $M=12$ also gives a straight line for the stall force
as a function of $\dE$ in the stiff limit, but neither consistent with
$\lim_{x\to\infty}\fscale{12}{x}=1$ nor with the value of $\dE_{12}$
from finite stiffness. This illustrates the effect of dense motor
subunits ($M>1/a$) in a stiff system.

The function $\fscale{M}{\km d^2/\dE}$ is a normalized stall force per
motor subunit. As is evident from Fig.\ \ref{scaleFig}(a), the force
production per motor subunit varies strongly with stiffness and also
with the number of subunits, $M$. Clearly, the interaction between the
motor subunits is important for the force generation. The stall force
is more complicated than a sum of contributions from the individual
parts. According to the arguments in Sec.\ \ref{qualitativeArguments},
one expects $\fscale{M}{x}$ to be an increasing function of $x$ for
$M<1/a$, since larger stiffness decreases the probability that the
unbound motor subunit binds to the wrong potential well. For $M>1/a$,
the stall force is expected to have a maximum as a function of
stiffness, reflecting the inset of the qualitatively different
behavior at high stiffness.  These expectations are consistent with
Fig.\ \ref{scaleFig}, where $\fscale{2}{x}$, $\fscale{3}{x}$,
$\fscale{5}{x}$, and $\fscale{6}{x}$ are monotonically increasing in
the simulated region, whereas $\fscale{12}{x}$ has a maximum around
$1.5$. To confirm that dense motor subunits give lower stall force
also for lower values of $M$, we performed simulations with $a=0.3$
and found that with $\dE=200$ pN nm and $M=5$, a maximum stall force
occurs for stiffness between 40 and 150 pN/nm (not shown), as
expected.

\subsection{Comparison with an individual motor subunit}
\begin{figure}[t]\begin{center}
\includegraphics[width=8.7cm]{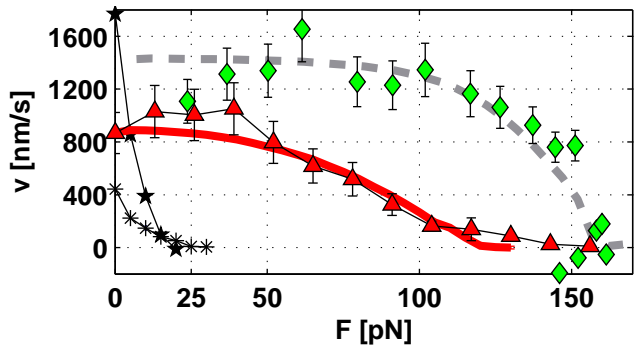}
\caption{(Color online) Force-velocity characteristics from simulation
and experiment.  Triangles: average experimental velocity from Fig.\
3(b) of Ref.\ \citep{maier04}.  Thick solid curve: $M=5$ model with
$\dE=200$ pN nm, $\km=25$ pN/nm, $\Fst=123$ pN, and $\lambda=1180$
s$^{-1}$.  Diamonds: a single retraction event from Ref.\
\citep{maier02}. Dashed curve: $M=5$ model with $\dE=200$ pN nm,
$\km=60$ pN/nm, $\Fst=160$ pN, and $\lambda=1800$ s$^{-1}$.  Isolated
motor subunits -- i.e., simple flashing ratchet models (black symbols)
are qualitatively different, as illustrated for $M=1$, $\dE=200$ pN
nm, $\lambda=2000$ s$^{-1}$, and $t_{\text{off}}=25$ ms ($\star$) or
$2.5$ ms ($\ast$).  }\label{VFfit}
\end{center}\end{figure}

To further highlight the effect of interactions and correlations, it
is interesting to compare the interacting ring model to an isolated
motor subunit. This is a single particle in a flashing ratchet
potential and a special case of the interacting model, with $M=1$ and
\mbox{$\km\to\infty$}. This single-particle flashing ratchet (SPR) has
been studied extensively \citep{reimann,astumian97} in different
versions, and its properties are qualitatively different from those of
the interacting model in several interesting respects. For an SPR in
this simple version, the mechanism to pull the particle to the next
potential well is not present. Instead, forward motion relies on
thermal noise to make the particle diffuse forward while in the
unbound state. This means that the probability for a forward step can
never exceed $1/2$ for a single chemical cycle even without applied
force. With nonzero applied force, the free diffusion is superimposed
on a backward motion with velocity
\mbox{$v_{\text{drift}}=-F/\gamma$}; hence, the velocity is
substantially reduced even at very low forces.  These features
conspire to make both the maximal velocity and the stall force depend
strongly on the friction constant, the reaction rate, and how much
time the particle spends unbound during a reaction cycle
\citep{astumian97,astumian94}.  Two examples of force-velocity
relations for SPR are shown in Fig.\ \ref{VFfit}, with 4 nm
periodicity, $\dE=160$ pN nm, $\lambda=2000$ $s^{-1}$, and damping as
above. The transitions between the bound and unbound states are
deterministic in these cases, and the time $t_{\text{off}}$ spent
unbound during each cycle was 2.5 ms and 25 ms.
\subsection{Application to a real motor system}
In Fig.\ \ref{VFfit} we compare model results with experimental data
on \bakterie{N. gonorrhoeae}. Several force-velocity relations
averaged over many retraction events, as well as two single events,
are preesented in Ref.\ \citep{maier02,maier04}.  The maximum forces
produced in the different events are distributed between 40 and 160
pN. There are several experimental factors that can make the maximum
measured force in a particular retraction event lower than the
intrinsic stall force \citep{maier02,maier04}. These factors include,
for example, breakage of the filament and are not included in the
model. Accordingly, the model should be compared to the data that
reach the highest forces. This leaves one single trajectory
\citep{maier02}, shown in Fig. \ref{VFfit}, and several average
curves. The average force-velocity data in Fig.\ \ref{VFfit} represent
an average of the data in Fig.\ 3 of \Ref\ \citep{maier04}.  Both
experimental curves give positive velocities up to about 160 pN, which
is in the upper tail of the maximum force distribution
\citep{maier04}, but the velocity at lower forces differs considerably
between the two curves.

To describe the PilT system, there are two natural choices for $M$:
namely, $M=5$, reflecting the five-fold symmetry of the filament, and
$M=6$, reflecting the six-fold symmetry of the PilT molecule. We
present simulation results for $M=5$, but $M=6$ is qualitatively no
different.

The model can describe both experimental curves to some extent. The
parameter values that describe the single event and the average data
(see Fig.\ \ref{VFfit}) differ in reaction $\lambda$ rate and
stiffness $\km$. There are two important deviations, which we discuss
next.

The average velocity falls off exponentially at high forces
\citep{maier04}, while the results of the model decay faster.  The
exponential decay can be described by an Arrhenius law for the rate
limiting step \citep{maier04}. The present model does not account for
this behavior.  

The single retraction event has a different decay at high forces and
agrees better with the characteristics of the model. However, the
single event shows an initial increase in velocity at low forces,
rather than a plateau. Such initial increases in velocity are also
obtained in some individual simulation runs, but disappear when the
average is computed. All the experimental data also suffer from a
possible systematic underestimation of the velocity near $F$=0
\cite{maier02,maier04}. Within the experimental accuracy, the general
trend is a constant velocity up to about $50$ pN
\cite{maier02,maier04}, which is consistent with the results of the
model.

\subsection{Deformations and elastic approximation}
From the results presented in Figs.\ \ref{scaleFig} and\ref{VFfit} and
Eq.\ \eqref{scaleEq}, it is possible to estimate the magnitude of the
deformations $y_i$ of the motor complex near the stall force. This is
useful, since the harmonic approximation for the confining force on
the motor subunits is questionable for large displacements.  At the
stall force, with $M-1$ subunits bound to the filament, the mean
displacement of a subunit can be estimated to $\Fst/\km(M-1)$. As seen
earlier, the stall force is less than $(M-1)\dE/d(1-a)$, which gives a
displacement less than $U/\km d(1-a)$. From Fig.\ \ref{scaleFig}, the
interesting and relevant regime with high normalized stall force has
$kd/U\gtrsim 1$ nm$^{-1}$. This gives a displacement less than $1$ nm
near the stall force. This is not excessively large compared to the
dimensions of the PilT ring, whose diameter is $11.5$ nm
\citep{forest04}.
\subsection{Role of order in the binding process}
We next examine the effect of the ad hoc assumption of a sequential
binding process on the results of the model. We compared the ordered
sequence with two less correlated binding schemes. For random order
with only one subunit free at the same time, the stall force is
essentially unchanged, but the velocity decreases with about
50\%. Alternatively, if the binding and unbinding events are assumed
to be independent for the different subunits, all subunits might
occasionally become unbound simultaneously, which releases the
filament from the motor. The resulting retraction events become highly
irregular and have low stall force and mean velocity. We conclude that
some degree of correlation between the motor subunits is necessary in
order for the model to simulate the experiments.
\section{Discussion and Conclusion}
We investigate a model for force generation in finite ensembles of
motor subunits interacting through an elastic backbone, which is
inspired by the pilus retraction machinery, the strongest molecular
motor reported so far.  The model is prototypical, rather than
realistic in detail, and offers a possible mechanism for generation of
large retraction forces.  It includes a ring of motor subunits
surrounding the pilus filament, following a suggestion in Ref.\
\citep{kaiser00}. Some parameters in the model can be roughly
estimated, and we explore parts of the remaining parameter space and
focus on generation of large forces.

We find that the stall force depends on the binding strength $\dE$
between motor and filament, the stiffness $\km$ of the motor complex,
the number of motor subunits $M$, and an asymmetry parameter $a$.  For
high enough stiffness we find qualitatively different properties
compared to the well-studied model of a single particle in a flashing
ratchet potential, which is the basic building block of our
model. This is not surprising, since the mechanisms that generate
motion are different in the two cases. The motion in the flashing
ratchet model is dependent on diffusion \citep{reimann}. Our model
also contains diffusive motion of the motor subunits, but diffusion is
not necessary for the motor to work \citep{dan03,klumpp01}.  The
dependence of the stall force on $\dE$, $\km$, and $M$ is well
parametrized in empirical scaling plots (Fig.\ \ref{scaleFig}). The
scaling ansatz in Eq.\ \eqref{scaleEq} relies on the presence of an
interaction between the motor subunits, and the stall force depends
nonlinearly on the number of motor subunits.

Low stiffness compared to binding strength has a strong destructive
effect on the force production. For a small number of motor subunits,
the stall force increases monotonically with increasing stiffness, but
when the motor subunits become dense enough there is a crossover to a
different regime, where a high stiffness instead has a destructive
effect on the stall force. For the asymmetry studied here, the
crossover occurs around $M=1/a$.  For $M=$ 5 or 6, corresponding to
the number of filament strands or PilT subunits, a system with $a>0.2$
would have the interesting property that the maximum stall force is
obtained for finite stiffness. If the binding strength between the
PilT subunits can be genetically engineered, it might be possible to
observe this effect in future experiments.

We compare results from the motor model for the filament retraction
force-velocity characteristics with experiments on \bakterie{N.\
gonorrhoeae} \citep{maier02,maier04}. Since the molecular details of
the retraction mechanism are unknown, it is unclear to what extent the
agreement we see reflects actual similarities between the model and
the real system.  More information would be useful for the
construction of more detailed models. Nevertheless, the model can
describe the general features of the experimental results -- i.e., the
plateau in the force-velocity relation at low applied forces and the
high stall force that is independent of reaction rate
\citep{maier02,maier04}.

For a quantitative comparison we select two different experimental
force-velocity relations that show large forces, one single event, and
one averaged curve, as shown in Fig.\ \ref{VFfit}. With different
parameters, the model gives a reasonable description of the average
data, as well as what looks like an atypical single event. The model
deviates from the average data at high forces, which indicates that
something is missing from the description.  However, the data might
include variations in cellular conditions that affect the average at
high forces. This is not accounted for in the simulations, where the
average is taken over thermal fluctuations with fixed parameters. The
single event is described better and does not suffer from such a
complication. In this case, the model is limited at high forces mainly
by thermally assisted transitions of motor subunits between potential
minima. Given the simplicity of the model, we find the agreement with
experiments encouaging.

Small ensembles of interaction motor systems have previously been
found to possess rich behavior without applied force
\citep{klumpp01,dan03,kolomeisky05}. As we have shown, this is true
also in the limit of high forces, which is a realistic experimental
situation.  Generation of strong forces in nanoscale devices is also
of technological interest, and it is tempting to speculate about the
possibility to realize a setup of interacting motors units pulling on
an artificial filament such as a carbon nano tube.
\begin{acknowledgments}
The authors thank Berenike Maier for valuable discussions and
comments. This work was supported by the Swedish Research Council
[Grants Nos.\ 2003-5001 (M.W.) 2001-6456, 2002-6240, and 2004-4831
(A.B.J.)], the Royal Institute of Technology, the G\"oran Gustafsson
Foundation, the Swedish Cancer Society, and Uppsala University.
\end{acknowledgments}

\bibliography{motorRefs}

\end{document}